\documentclass[summary]{gass2017}
\usepackage{astrojournals}
\usepackage{amsmath}
\usepackage{amssymb}
\usepackage{natbib}
\usepackage{subcaption}
\providecommand{\HI}{{H\,\textsc{i}}}

\title{Self-calibration of highly-redundant low-frequency arrays --
  initial results with HERA}

\author{Bojan Nikolic*\affref{ref1} \&
  Chris Carilli\affref{ref2}\affref{ref1}, on behalf of the HERA collaboration}

\affiliation{%
  \aff{ref1}{Astrophysics Group,  Cavendish Laboratory, University of Cambridge, UK}
  \aff{ref2}{National Radio Astronomy Observatory, Socorro, NM, USA}
}

\begin{document}

\maketitle

\begin{abstract}
  HERA is a highly-redundant transit interferometer with
  14\,m-diameter parabolic dish elements. We exploit the fact that the
  Galactic centre transits through the main beam of the telescope to
  attempt a conventional self-calibration approach to imaging and
  calibration. The Galactic centre provides a bright source which, we
  show, can be approximated as a point source sufficiently well to
  initialise the self-calibration loop and derive initial delays and
  antenna frequency-independent phases.  Subsequent iteration using a
  more complex sky model derived from the data itself then converges
  to a reasonable bandpass calibration. The calibration solutions have
  good stability properties. We show therefore that the conventional
  self-calibration is a feasible parallel approach in addition to the
  redundant calibration already planned for HERA. The conventional
  imaging and calibration is useful as a cross-check to the
  alternatives being pursued in the HERA project, as a way of
  quantifying the performance of the hardware on the ground (and
  potentially identifying problems) and as a route to imaging and
  removing brighter continuum sources before power spectrum analysis.
\end{abstract}

\section{Introduction}

The Hydrogen Epoch of Reionization Array
\cite[HERA,][]{2016arXiv160607473D} is an experiment to measure the
large-angular-scale redshifted \HI\ 21\,cm line signal from the Epoch
of Reionisation (EoR, $ 12\lesssim z \lesssim 6$).  HERA is located at
the Karoo Radio Astronomy Reserve in South Africa and as a scientific
pathfinder for the SKA situated on an SKA site it has been designated
as a SKA Precursor instrument.

The full HERA instrument will consist of up to 350 elements each of
which is a 14\,m-diameter parabolic antenna fixed in the transit
direction. Of these, 320 elements will be arranged in a dense-packed
hexagonal configuration and 30 will be arranged as outrigger elements
on baselines up to 800\,m.  The instrument is being deployed in
stages: in this paper we present measurements with the first 19
antennas; at the time of writing there are 37 antennas on the site and
it is expected by the end of the 2017 calendar year there will be 128
operational antennas on the site.

HERA is designed to have highly redundant baselines to provide high
sensitivity for the delay-spectrum analysis strategy
\citep{2012ApJ...753...81P} as well as to enable redundant baseline
calibration \citep{2010MNRAS.408.1029L}. These two techniques form the
primary strategy for scientific data analysis for HERA.  In this
summary paper we show initial results of calibrating HERA with the
conventional imaging and self-calibration approach with a view of
investigating if this is a feasible parallel data reduction route.

\section{Observations}

HERA observes continuously during the night hours. The observations we
selected for analysis were made during June 2016 with 19 elements at
the Karoo site. All of the elements were in a closed-packed hexagonal
configuration.  Observations from this time were chosen because the
Galactic centre transits through the telescope main beam during this
time. The observations consist of 1024 channels covering 100\,MHz to
200\,MHz and have an integration time of 10.7\,s.

\section{Calibration}

The strategy we employed was to derive the delays and initial
frequency-averaged gain phase solutions from an initial sky-model
consisting simply of a point source at the Galactic centre, then make
an initial image of the sky which then was used for solving for the
full-resolution band-pass calibration.  All of the processing was done
in the NRAO CASA package.

\begin{figure}
\begin{subfigure}[t]{\linewidth}
  \includegraphics[width=\linewidth,trim=50 400 050 70,clip]{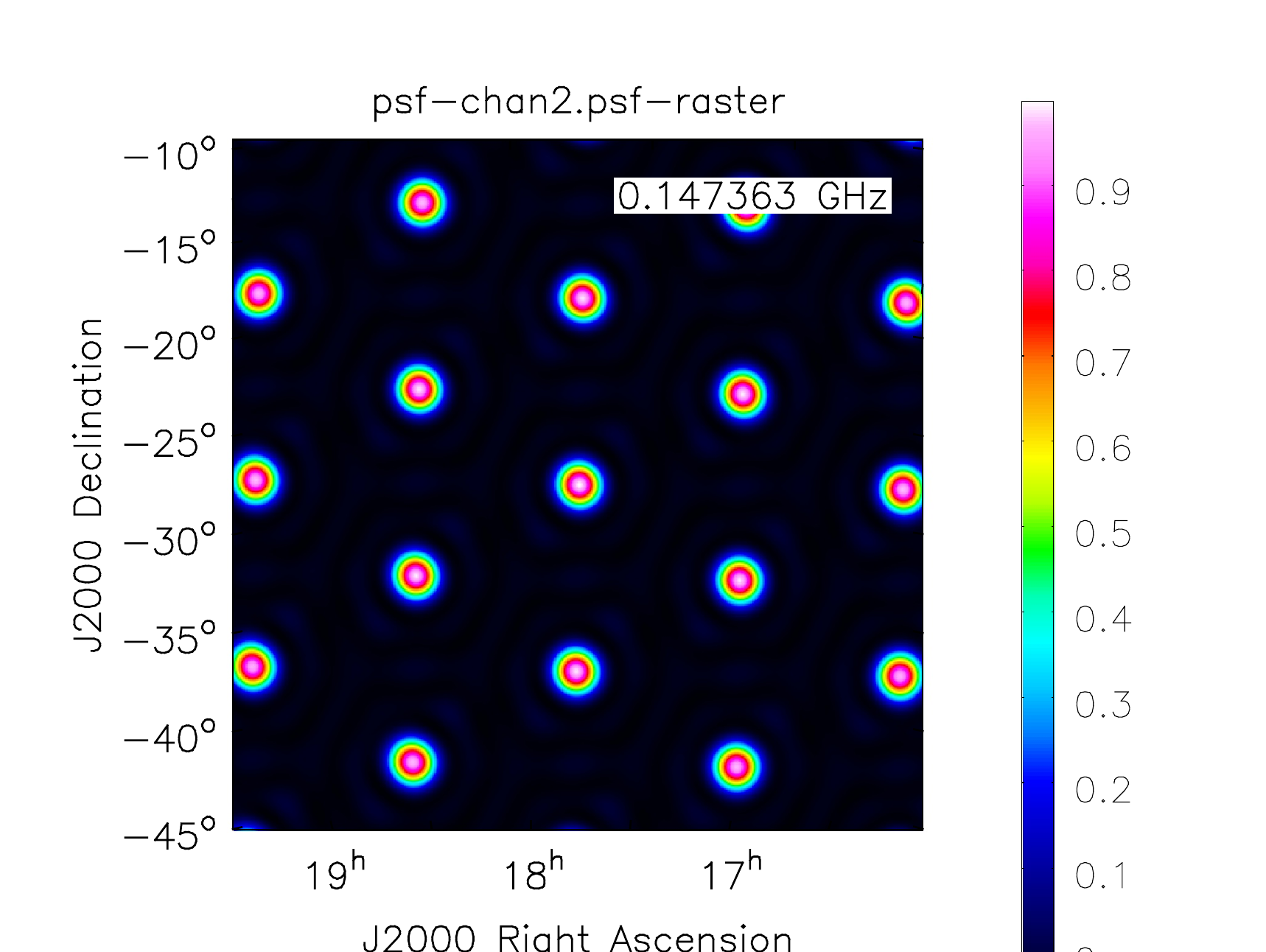}
  \caption{PSF for a single channel (close to the middle of the
    observing band) map}
\end{subfigure}
\begin{subfigure}[t]{\linewidth}
  \includegraphics[width=\linewidth,trim=50 400 050 70,clip]{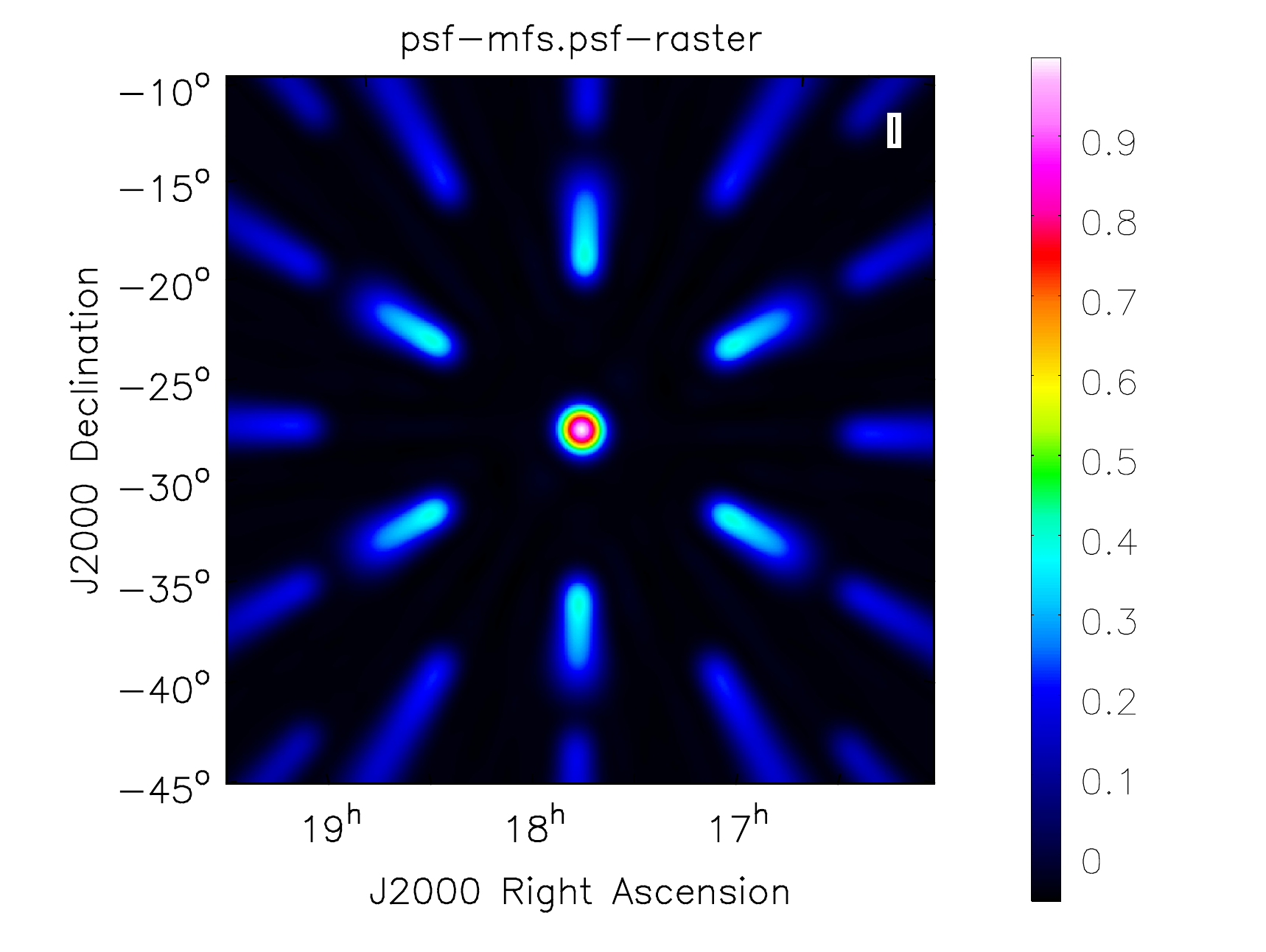}
  \caption{PSF for a multi-frequency synthesis map of the whole
    observed band (100\,MHz -- 200\,MHz)}
\end{subfigure}
\caption{Illustration of the contrast between (a) a single-channel PSF
  and (b) multi-frequency synthesis PSF. The PSF was computed from a
  synthesis of about 20 minutes of data.}
\label{fig:psf}
\end{figure}

The calibrations steps for a dataset with the Galactic centre in the
main beam were:
\begin{enumerate}
\item Carry out initial visual flagging of bad antennas (2) and
  channels affected with RFI
\item Insert a (flat-spectrum) continuum point source model for the
  Galactic centre
\item Use this point-source model for solve for the antenna delays and
  the mean antenna phase  (i.e., phase averaged over the whole band)
\item Apply the delay and phase corrections and make an image of the
  sky using the MultiFrequency Synthesis (MFS) mode of CASA
  \citep{2011A&A...532A..71R}. The CLEAN components were constrained
  to be inside a region corresponding to the nominal primary beam of
  HERA
\item Using the CLEAN model derived in previous step, solve for the
  bandpass amplitude and phase at full spectral resolution
\item Apply the bandpass solution and repeat the imaging step for a
  visual check. Optionally repeat the cycle of bandpass calibration
  and imaging
\end{enumerate}
We found that use of MFS is essential for this process to converge
both to decrease the point spread functions sidelobes (which are very
pronounced due to the highly redundant configuration of HERA, see
Figure~\ref{fig:psf}) and to improve the signal to noise ratio.

\section{Results}

\begin{figure}
\begin{subfigure}[t]{\linewidth}  
  \includegraphics[width=\linewidth]{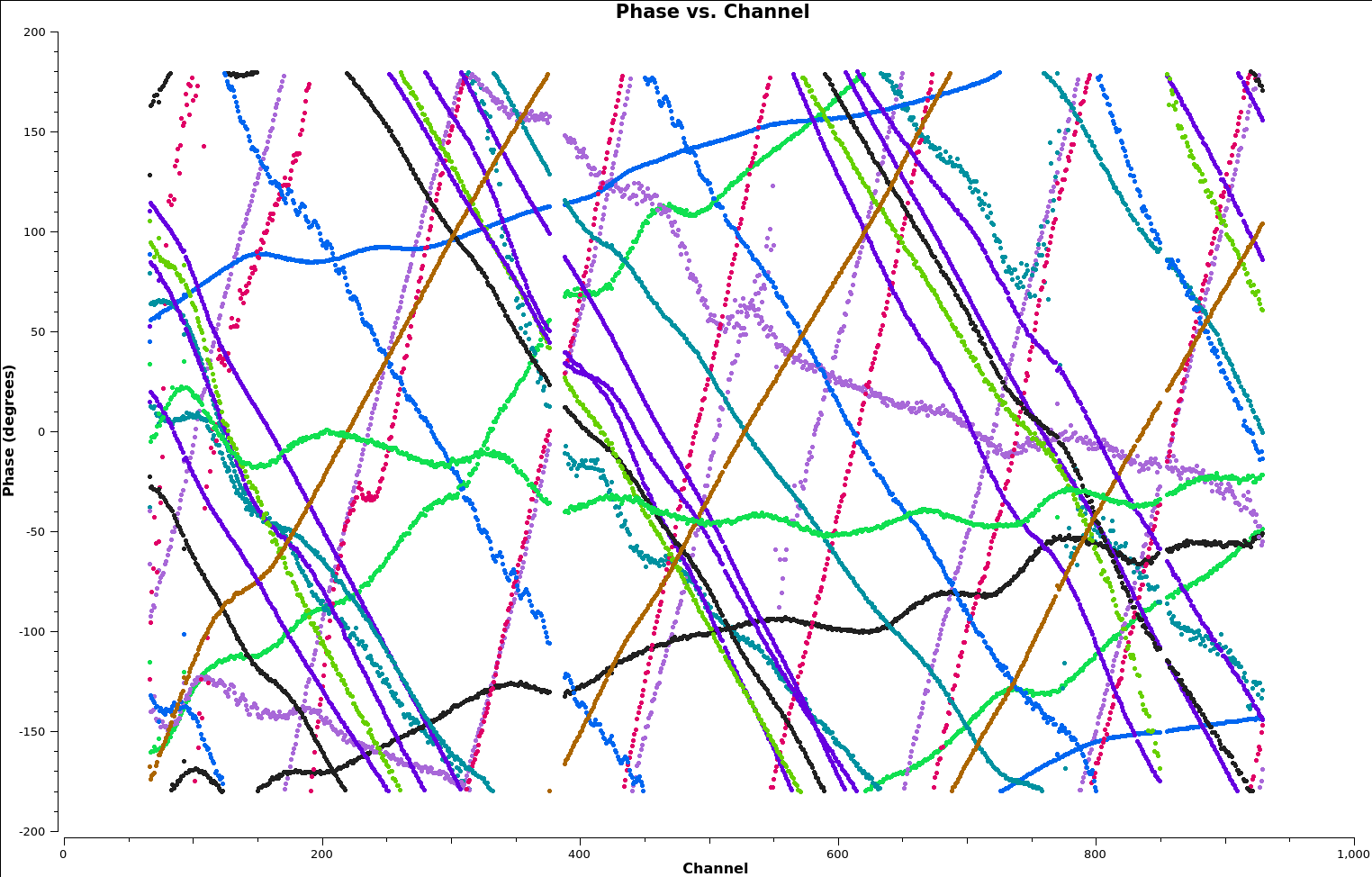}
  \caption{Fringes for all baselines to one of the HERA antennas
    before any calibration when the Galactic centre transits through
    the beam}
\end{subfigure}
\begin{subfigure}[t]{\linewidth}  
  \includegraphics[width=\linewidth]{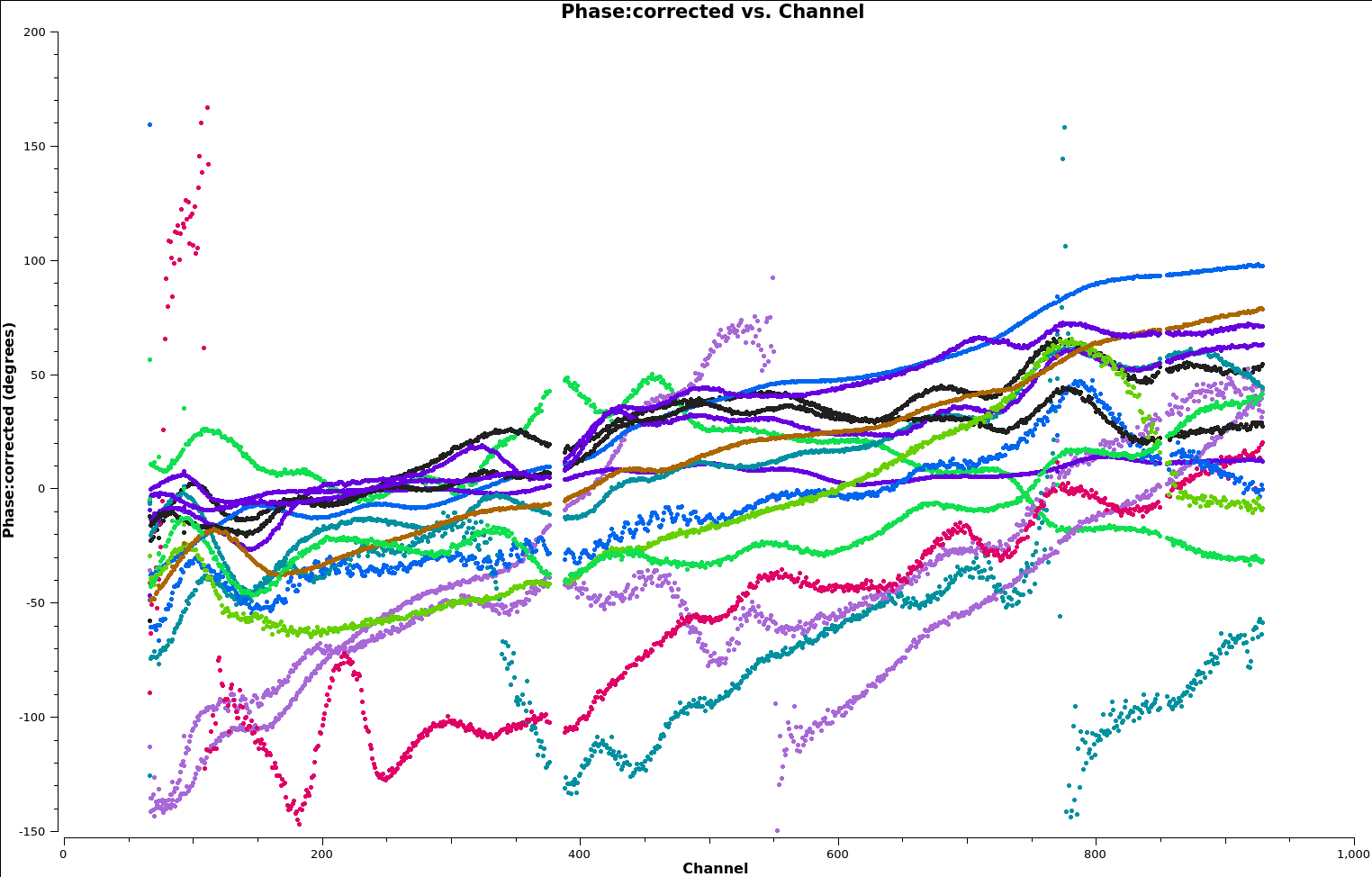}
  \caption{The same fringes as in (a) but after the initial delay and
    frequency-independent antenna phase calibration}
\end{subfigure}
\begin{subfigure}[t]{\linewidth}  
  \includegraphics[width=\linewidth]{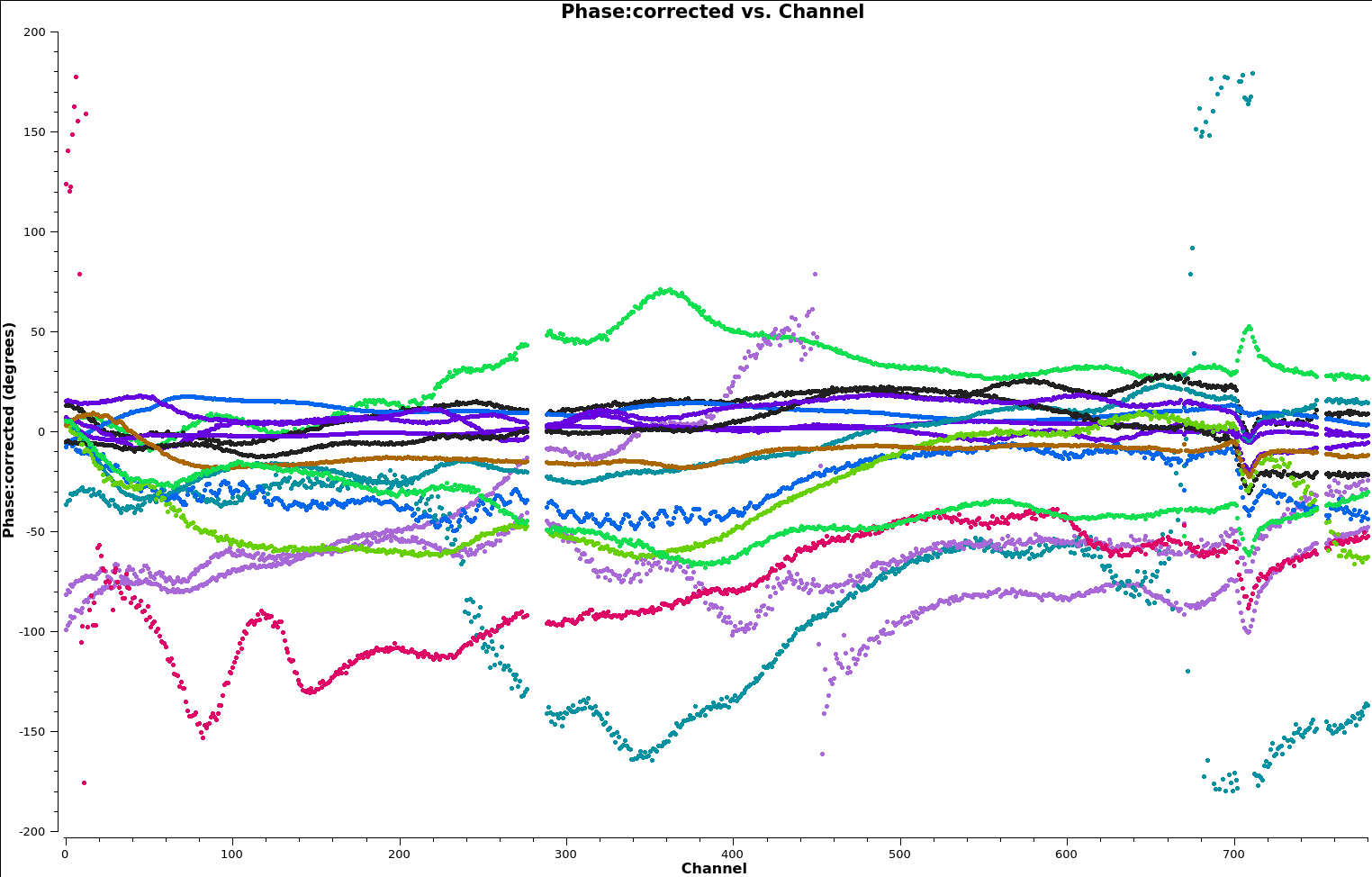}
  \caption{The same fringes as in (b) but after additionally
    correcting the band-pass response using the calibration based on
    the sky model derived from the data themselves}
\end{subfigure}
\caption{Illustration of the effect of calibration on the observed
  fringes when the Galactic centre is transiting.}
\label{fig:cal}
\end{figure}

\begin{figure}
\includegraphics[width=\linewidth,trim=100 450 050 150]{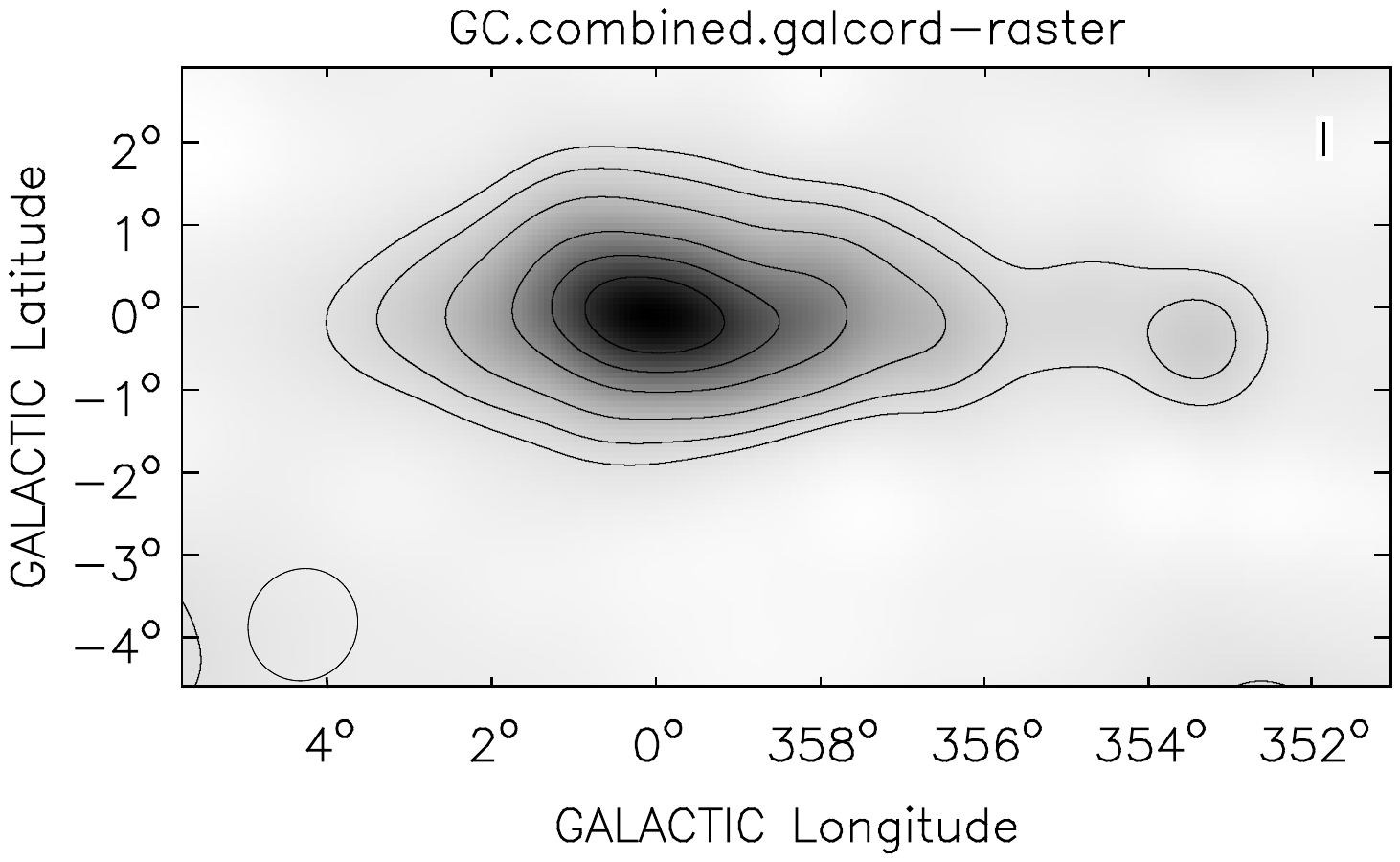}
\caption{Final image of the Galactic Centre with HERA-19 using 20
  minutes of data. The whole observing band was imaged as a continuum
  using multi-frequency synthesis. Contours are shown at
  $0.05, 0.1, 0.2, 0.4, 0.6, 0.8\times$peak. The data have not been
  placed on an absolute flux scale.}
\label{fig:GCimage}
\end{figure}

For the present analysis we've concentrated on datasets which have the
Galactic centre transiting through the main beam. We find that the
above procedure results in good calibration solutions as judged by:
\begin{enumerate}
\item Feedback from CASA about the signal-to-noise ratio of the
  calibration solutions (few solutions are flagged due to insufficient
  S/N)
\item The fringe pattern after calibration, shown in
  Figure~\ref{fig:cal}, is consistent with observations of a
  point-source dominated region of the sky
\item The resulting image of the sky, shown in
  Figure~\ref{fig:GCimage}, is consistent with known structure in the
  Galactic centre
\end{enumerate}

\begin{figure}
 \begin{subfigure}[t]{\linewidth}  
\includegraphics[width=\linewidth]{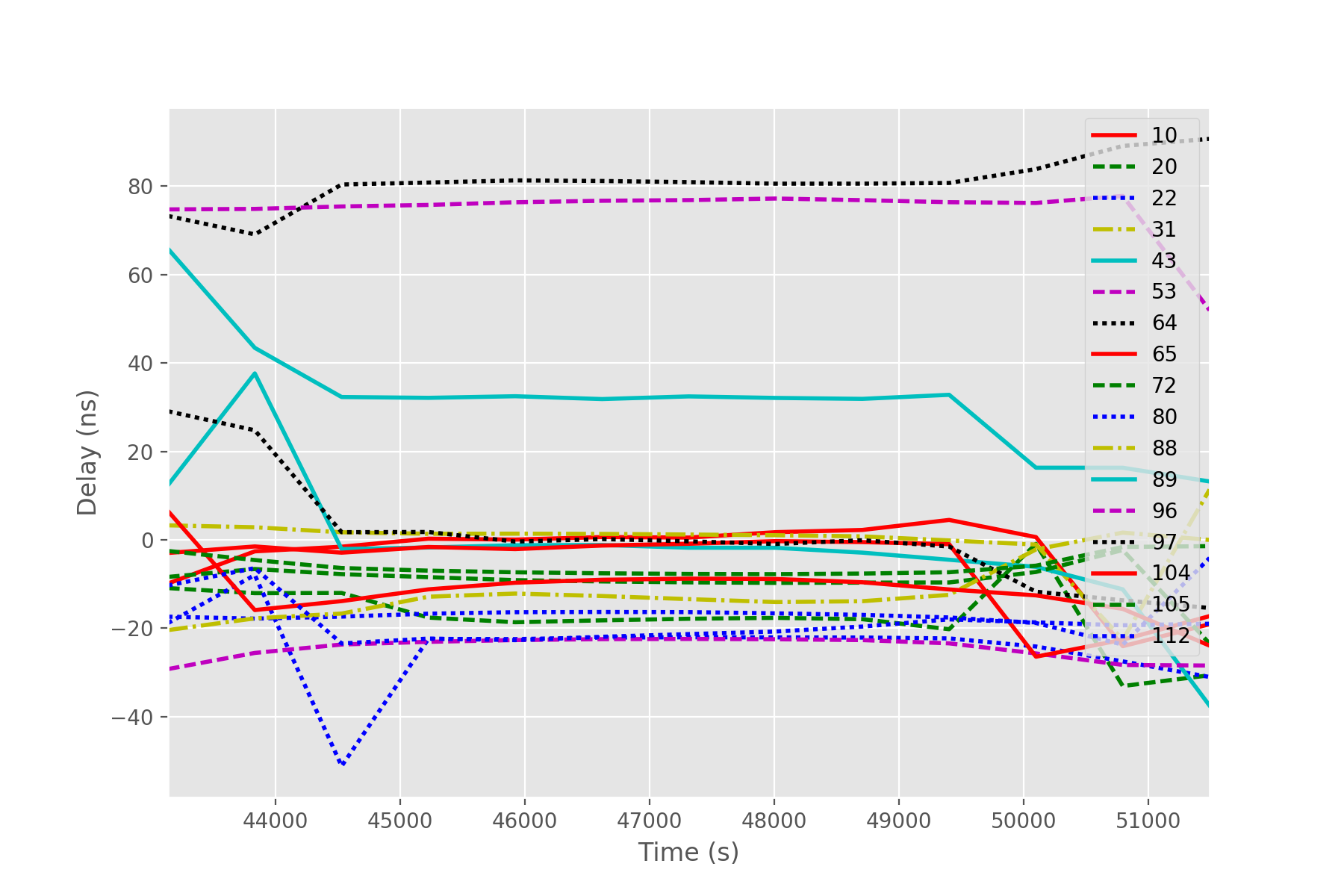}
\caption{Antenna delay solutions derived from a sequence of 13
  observations around the time of the transit of the Galactic
  centre. The fluctuations at the beginning and end are due to the
  Galactic centre being outside the main beam of the telescope at
  these times.}
\end{subfigure}
\begin{subfigure}[t]{\linewidth}  
\includegraphics[width=\linewidth]{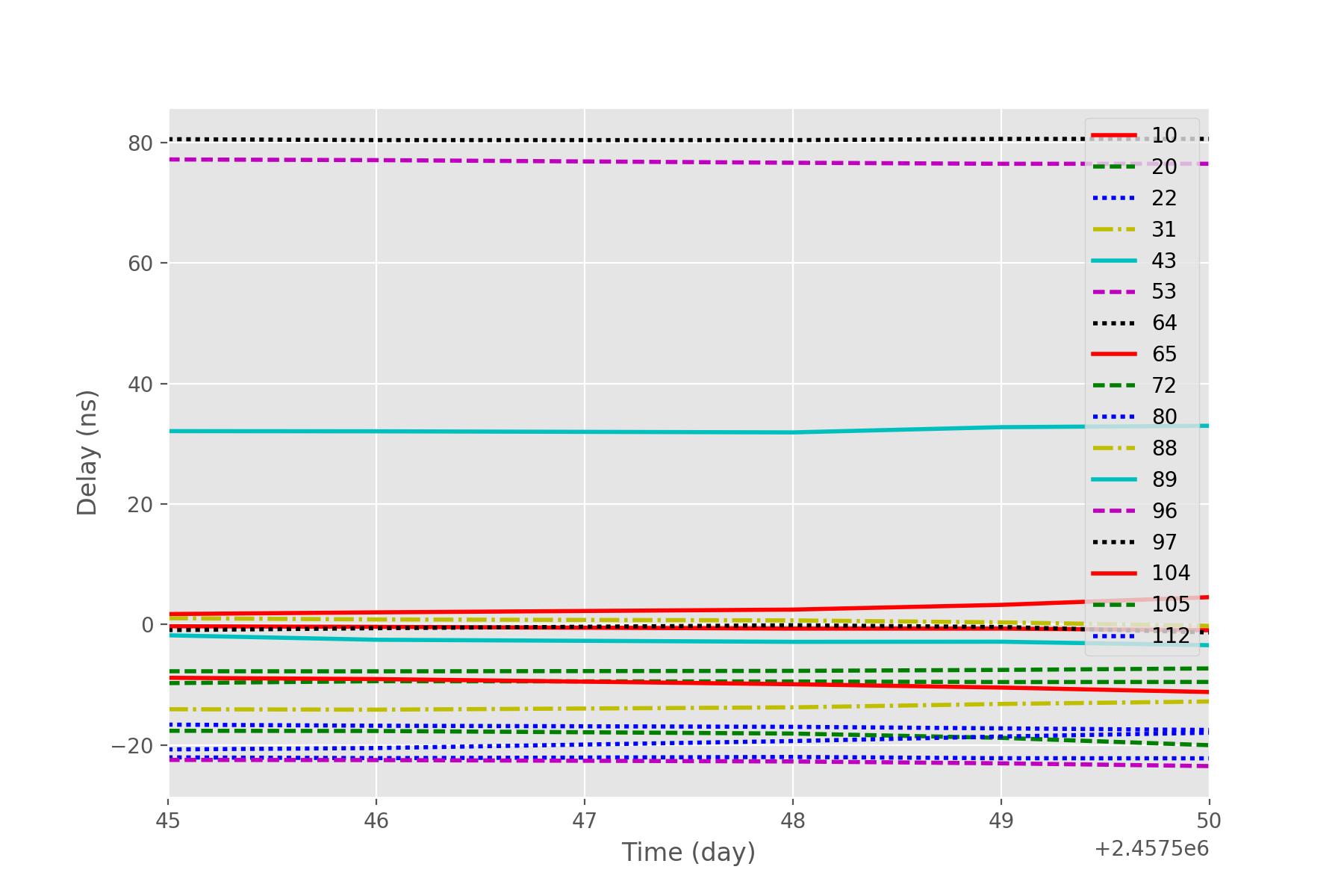}
\caption{Antenna delay solutions derived from a sequence
  10\,minute-long observations around the time of the transit of the
  Galactic centre on five consecutive days.}
\end{subfigure}
\caption{Illustration of the stability of the delay solutions made
  with observations of the Galactic centre during a day (top panel)
  and across different days (lower panel).}
\label{fig:delays}
\end{figure}

We have investigated the stability of the delay solutions when
analysing datasets just before and just after the transit of the
Galactic centre. The results are shown in Figure~\ref{fig:delays},
where it can be seen that the good solutions are found for an
hour-long period around the transit of the Galactic centres. During
this time the delays are highly stable. As expected, when the Galactic
centre is outside the main beam (the first few and last few points in
Figure~\ref{fig:delays}) it is not possible to solve for the delays
using the Galactic centre point source model.

In the lower panel of Figure~\ref{fig:delays} we show the variation of
the delay solutions for observations made at same local time over five
consecutive days. The delays can be seen to have good stability over
this period.

\section{Discussion}

We show that it possible to do a reliable calibration of a highly
redundant array such HERA using the conventional self-calibration
technique. Our calibration is initialised using a point source model
for the Galactic centre and is then refined by making a map from the
data itself and using that to solve for the frequency dependent
structure of the antenna gains. The stability of the solutions is
sufficient that it can be transferred in time to other fields,
allowing these fields to be imaged and potentially making these other
fields also useful for calibration.

This conventional imaging and calibration is useful as a cross-check
to the redundant-calibration alternative also being pursued in the
HERA project, as a way of quantifying the performance of the hardware
on the ground (and potentially identifying problems) and as a route to
imaging and removing brighter continuum sources before power spectrum
analysis.

\section{Acknowledgements}

The work is here is based on observations by the HERA
collaboration. We gratefully acknowledge the contribution of the whole
HERA team in acquiring the data as well as the feedback on an early
version of this analysis.

\bibliographystyle{egu}
\bibliography{hera}

\end{document}